\documentclass[aps,floatfix]{revtex4}
\usepackage{graphicx}
\usepackage{epsfig}

\newcommand{\be}{\begin{eqnarray}}
\newcommand{\ee}{\end{eqnarray}}
\newcommand\del{\partial}
\def\conj#1{{{#1}^{*}}}

\newcommand\hatmu{\hat{\mu}}

\newcommand{\mat}{\left ( \begin{array}{cc}}
\newcommand{\emat}{\end{array} \right )}
\newcommand{\matf}{\left ( \begin{array}{cccc}}
\newcommand{\ematf}{\end{array} \right )}            
\newcommand{\matt}{\left \begin{array}{ccc}}
\newcommand{\ematt}{\end{array} \right )}
\newcommand{\vect}{\left ( \begin{array}{c}}
\newcommand{\evect}{\end{array} \right )}

\newcommand{\nn}{\nonumber } 

\def\d{\partial}
\newcommand{\tr}{{\rm Tr}}

\begin{document}
\setcounter{page}{0}
\thispagestyle{empty}



\begin{flushright}
\end{flushright}

\title{QCD with Bosonic Quarks at Nonzero Chemical Potential}

\author{K. Splittorff}
\affiliation{The Niels Bohr Institute, Blegdamsvej 17, DK-2100, Copenhagen {\O}, Denmark}
\author{J.J.M. Verbaarschot}
\affiliation{Department of Physics and Astronomy, SUNY, Stony Brook,
 New York 11794, USA}

\date   {\today}
\begin  {abstract}
We formulate the low energy limit of QCD like partition functions
with bosonic quarks at nonzero chemical potential. 
The partition functions are evaluated 
in the parameter domain that is dominated by the zero momentum modes 
of the Goldstone fields.
We find that partition functions with bosonic quarks differ structurally 
from partition functions with fermionic quarks.
Contrary to the theory with one fermionic flavor, where the partition function
in this domain does not depend on the chemical potential, a
phase transition takes place in the theory with one bosonic flavor 
when the chemical potential is equal to $m_\pi/2$. 
For a pair of conjugate bosonic
flavors the partition function shows no phase transition, whereas
the fermionic counterpart has a phase transition at $\mu = m_\pi/2$. 
The difference between the bosonic theories and the fermionic ones
originates from the convergence requirements of bosonic integrals
resulting in a noncompact Goldstone manifold and a covariant derivative
with the commutator replaced by an anti-commutator.  
For one bosonic flavor the partition function is evaluated using a
random matrix representation.
\end {abstract}

\maketitle



\section{Introduction}

Bosonic quarks appear in QCD whenever the weight of the partition
function includes an inverse determinant of the Dirac operator. Ratios
of determinants
occur frequently in analytical approaches to QCD. Inverse 
determinants are for example used to quench a flavor from the theory 
\cite{SUSY,replica}. This form of  
quenching is useful in describing quenched and partially quenched
\cite{PQ,replicaPQ} lattice results and is an essential ingredient when 
computing the spectral properties of the QCD Dirac operator. 
In particular, if we consider the spectral correlation functions of the Dirac
operator in the microscopic limit 
\cite{LS,SV}, the theories with bosonic flavors are
not only a part of the calculation, they are also part of the result
\cite{SplitVerb1,SplitVerb2,SplitVerb3}.   

The microscopic limit is an extreme version of the chiral limit where the
sea and valence quark masses times the volume are kept fixed as the volume is 
taken to infinity. In this limit the zero modes of
the Goldstone bosons associated with chiral symmetry breaking
dominate the low energy partition function which reduces to a group integral
uniquely determined by the pattern of chiral symmetry breaking \cite{GLeps}. 
Sometimes this limit is referred to as the {\sl $\epsilon$-limit} but since
the term {\sl microscopic limit} appeared earlier in the literature (see for
example \cite{Efetov,VZ}) we prefer to use this term. 
For a review of QCD in the microscopic
limit see \cite{Tilo-Jac}.

In this paper we will consider the low energy limit of QCD 
at nonzero chemical potential. For fermionic quark flavors the baryon chemical
potential does not affect the low energy effective 
partition function at zero temperature. The reason is that the lightest 
degrees of freedom (the pions) have zero baryon charge. Because the fermion
determinant is complex, QCD at nonzero baryon chemical potential cannot be 
simulated directly on the lattice by probabilistic methods. 
That is why sometimes a theory is considered where the fermion determinant
is replaced by its absolute value. For an even number of flavors this
corresponds to the product of a fermion determinant and its complex
conjugate in the weight of the partition function. The low energy partition 
function for QCD with flavors and conjugate flavors depends on the chemical 
potential \cite{misha}.
The conjugate fermionic flavors correspond to ordinary fermionic flavors with
the opposite sign of the chemical potential \cite{AKW}. A theory with one
fermionic flavor and a 
conjugate fermionic flavor is therefore identical to  a theory 
with two fermionic flavors
at nonzero isospin chemical potential. Since the pions have nonzero
isospin charge this low energy effective partition function depends on the
chemical potential. In particular, a phase transition to a Bose-Einstein
condensed phase takes place at zero temperature when the isospin chemical
potential reaches $m_\pi/2$, see \cite{misha,KST,KSTVZ,eff,lattice}.    

In this paper we examine the properties of the  low energy 
QCD partition function with bosonic
quarks at nonzero chemical potential. We find that the bosonic theories
behave completely different 
from their fermionic counterparts. The bosonic partition functions 
depend on the chemical potential even in the absence of conjugate bosonic
flavors. By explicit computation 
of the partition function with one bosonic 
flavor in the microscopic limit we find that this theory has a phase transition
when the chemical potential, $\mu$, reaches $m_\pi/2$. On the contrary, in
the theory 
with one conjugate pair of bosonic quarks we find that the free energy as
well as its derivative are continuous at $\mu=m_\pi/2$ and zero
temperature. 
The difference in the phase structure between the bosonic and the fermionic
theories has its origin in the convergence requirements of bosonic integrals
which forces us to rewrite the  determinants in the partition function
 as the determinant
of a Hermitian matrix (also known as hermitization 
\cite{Feinberg,Janik,Efetov-direct}). This procedure leads to a noncompact
Goldstone manifold and, as will be shown here, a covariant derivative where
the commutator is replaced by an anti-commutator.
A comparison of bosonic and fermionic partition functions is given in
table \ref{table:summary}.

In order to compute the partition function with one bosonic flavor we 
make use of the random matrix representation. All other partition functions
discussed here have been derived directly from the low energy effective 
theory.

\vspace{2mm}

The outline of this paper is as follows. We first discuss in general terms
the low energy physics of QCD with bosonic quarks at nonzero chemical
potential. As a warm up exercise we review  
results for fermionic quarks at nonzero chemical potential. 
We then turn to the theories with conjugate
pairs of quarks.  
In section \ref{sec:rmt} we present a random matrix model and perform the 
calculation of the partition function with one bosonic flavor in 
the microscopic limit. Taking the thermodynamic limit of this result then
allows us to examine the phase diagram of QCD with one bosonic quark.  

\section{General Discussion}
\label{sec:gen}

In this section we give a general discussion of QCD-like partition
functions at nonzero chemical potential. The Euclidean QCD partition function
at nonzero chemical potential for $N_f$ flavors with mass $m$ 
is given by
\be
Z^{N_f}(m;\mu) = \langle {\det}^{N_f}( D + \mu \gamma_0 +m) \rangle.
\label{znf}
\ee
Here and below $\langle\ldots\rangle$ denotes the average with respect to the
Yang-Mills action. 
If $N_f>0$ the quarks are fermions while for negative $N_f$  they are 
interpreted as 
bosons. With fermionic quarks at zero temperature this partition function 
is independent of the chemical
potential for $\mu < m_N/3$, which immediately follows from the
definition of the grand canonical partition function as an average over
Boltzmann factors and fugacities. 
In terms of the representation 
(\ref{znf}), the $\mu$-independence of the free energy in the thermodynamic
limit can be understood 
from the gauge transformation
\be
D + \mu \gamma_0 +m = e^{-\mu t} [ D+ m ]e^{\mu t},
\ee
so that the factor $\exp(\pm\mu t)$ can be absorbed in 
the boundary conditions of the fermionic fields. In the phase
that is not sensitive to the boundaries, the partition function does not
depend on $\mu$. A nonzero baryon density is obtained by baryons winding
around the torus in the time direction. In this phase the boundary 
conditions are important and the partition function becomes $\mu$-dependent.

The second partition function we consider is the phase quenched partition
function (the superscript $n$ counts the pairs of conjugate fermionic,
$n>0$, or bosonic, $n<0$, determinants) 
\be
Z^{n=1}(m,m^*;\mu) = \langle \det (D +\mu \gamma_0 +m ) 
\det(D +\mu\gamma_0 +m)^\dagger\rangle=
\langle \det (D +\mu \gamma_0 +m ) 
\det(-D +\mu\gamma_0 +m^*)\rangle,
\ee 
so that $\mu$ can be interpreted as the isospin chemical potential \cite{AKW}.
Also in this case the chemical potential can be gauged into the
boundary conditions, this time by a gauge transformation in isospin
space. 
For $\mu <
m_\pi/2$ the partition function is $\mu$-independent at zero temperature. 
Again, this also follows from the zero temperature limit of the Boltzmann
factors. 

Next, let us consider the bosonic partition function
\be
Z^{n=-1}(m,m^*;\mu) = \langle \frac{1}{\det (D +\mu \gamma_0 +m ) 
\det(D +\mu\gamma_0 +m)^\dagger}\rangle=
\langle \frac{1}{\det (D +\mu \gamma_0 +m) 
\det(-D +\mu\gamma_0 +m^*)}\rangle.
\label{zn=-1}
\ee
Because of the nonhermiticity, the inverse determinants {\it cannot} be
written directly as a convergent bosonic integral. However, this can be
achieved by introducing the infinitesimal regulator $\epsilon$
\be
Z^{n=-1}(m,m^*;\mu) = \left \langle\det \mat \epsilon & D + \mu \gamma_0 +m \\
                 -D + \mu\gamma_0+m^* & \epsilon \emat^{-1}  \right \rangle .
\label{zphq}
\ee 
The parameter $\epsilon$ 
may be regarded as source for pions composed of bosonic anti-quarks and 
conjugate bosonic quarks.
Expressing the partition function as an integral over the eigenvalues
of the Hermitian matrix in (\ref{zphq}) one can easily convince oneself
that the partition function diverges logarithmically in $\epsilon$.
In \cite{SplitVerb2} this was shown explicitly by performing the 
integrals in the low-energy limit of this partition function.
Because of the $\epsilon$-term it is not possible to gauge away the
$\mu$-dependence of the partition function, and we do not expect to find
a phase where the partition function is $\mu$-independent. Indeed, we
show below that the partition function (\ref{zphq}) is $\mu$-dependent 
for all values of $\mu$. Therefore the pion condensate, which has
$\epsilon$ as source term, is nonzero for all values of $\mu$.
In particular this implies that there must be a massless mode in the
spectrum. We identify this mode explicitly below.

Finally, we discuss the partition function
\be
Z^{N_f =-1}(m;\mu) = \left \langle \frac 1{\det (D+\mu\gamma_0 +m) }\right \rangle.
\ee
It is not possible to write this partition function as a convergent 
bosonic quark integral. To properly define the partition 
function we have to rewrite it as
\be
\label{ZNf-1structure}
Z^{N_f =-1}(m;\mu) = \left \langle \frac {\det(-D+\mu \gamma_0+m^*)}
{\det (D+\mu\gamma_0 +m) \det (-D+\mu\gamma_0 +m^*) }\right \rangle
\ee
so that the inverse determinant  can be represented as a bosonic integral 
after it has been  regularized
as in (\ref{zphq}). 
The particle content of this partition function is one conjugate
fermionic quark, one bosonic quark, and one conjugate bosonic quark.  
Because of the extra determinant in
the numerator, the limit $\epsilon \to 0$ is finite in this case, i.e.~the
leading term is of order $\epsilon^0$. In the partition function the gauge
symmetry therefore allows us to transform the $\mu$ dependence to the 
boundaries allowing for a phase transition to occur.  
As we will show below a phase exists where this partition function does 
not depend
on $\mu$. For $\mu>m_\pi/2$ this phase gives way to a $\mu$ dependent phase.

A summary of the different partition functions 
discussed in this section is given in the table below.  

\begin{table*}[htb]
\begin{ruledtabular}
\begin{tabular}{c|c|c}
& & \\
{\large \textrm{Theory}} & {\large \textrm{ Number of Charged Goldstone 
$ \ \ \ $}} 
&{\large \textrm{ Critical Chemical Potential  $ \ \ \ $}}\\
& {\large \textrm{  
Modes for $\mu < \mu_c$}}  &  \\ 
 & & \\
\hline
&&\\
{\large $\langle\det(D+\mu\gamma_0+m)\rangle$} & {\large 0} 
&\bf \large  $\mu_c = \frac 13 m_N $\\
&&\\
\hline
&&\\ 
{\large $\langle|\det(D+\mu\gamma_0+m)|^2\rangle \ \ \ $} & {\large  
$2 $}
 & {\bf \large $\mu_c = \frac 12 m_\pi$} \\
&&\\
\hline
&&\\
{\large $\langle\frac{1}{\det(D+\mu\gamma_0+m)}\rangle$} & {\large $4$} 
& {\bf \large $\mu_c = \frac 12 m_\pi$} \\
&&\\
\hline
&&\\
{\large $\langle\frac{1}{|\det(D+\mu\gamma_0+m)|^2}\rangle$} & {\large 
non applicable}
 & {\bf \large $\mu_c = 0$} \\
&&\\
\end{tabular}
\end{ruledtabular}
\caption{\label{table:summary}
Summary of properties of low energy QCD at nonzero chemical potential and
zero temperature.}
\end{table*}
 
Notice that the regularization enters because the chemical potential breaks the
hermiticity properties of the Dirac operator. If we where to consider an {\sl
  imaginary} chemical potential, the Dirac spectrum would remain on the
imaginary axis and no regularization of the bosonic theories would be
required \cite{DHSS}. 
\section{Fermionic Partition Functions}
\label{sec:ferm}

In the microscopic limit the zero momentum modes of the pions dominate the low
energy effective partition function of QCD in the sense that the nonzero
momentum modes factorize from the partition function leaving us with 
a group integral over the zero momentum modes. 
This integral is uniquely determined by the
pattern of chiral symmetry breaking, and in the case of bosonic quarks,
by the convergence of the integrals.
Before turning to the theories with bosonic quarks, 
as a warm-up exercise, we recall results obtained 
for fermionic quarks. The emergence of the phase structure is discussed 
as well.

\vspace{2mm}

With fermionic quarks the partition function is automatically convergent and
we need only worry about the symmetries. The Lagrangian is determined
by local gauge invariance in isospin space \cite{KST}. 
For $N_f+n$ ordinary fermionic quarks and $n$ conjugate fermionic quarks it is
given by \cite{TV} 
\be
{\cal L} = 
\frac 14 F_\pi^2 {\rm Tr} \nabla_\nu U \nabla_\nu U^{-1} - \frac \Sigma 2 
{\rm Tr} M(U+U^{-1})
\label{lferm}
\ee
with $U \in SU(N_f +2n)$, and
the quark mass matrix is given by 
$M$=diag$(m_1,\ldots,m_{N_f},\{z\}_n,\{z^*\}_n)$.  
The charge matrix $B$ is the  diagonal matrix 
$B = {\rm diag}(1,\cdots, 1_{N_f+n}, -1,\cdots, -1_n)$.
The chiral Lagrangian is parameterized by two low energy constants,
the chiral condensate, $\Sigma$, and the pion decay constant $F_\pi$.
The covariant derivatives are defined by
\be
\nabla_\nu U = \del_\nu U + \mu\delta_{\nu \, 0}[U,B], \qquad 
\nabla_\nu U^{-1} = \del_\nu U^{-1} + \mu\delta_{\nu\, 0}[U^{-1},B].
\label{covferm}
\ee

In the microscopic limit the nonzero momentum modes factorize from
the partition function \cite{GLeps} so that the zero momentum part
of the partition function in the sector of zero topological charge 
is given by \footnote{Without 
loss of generality we consider the
  partition function in the trivial topological sector.}   
\be
Z^{N_f,n}(\{m_f\},z,z^*;\mu) = 
\int_{U \in U(N_f+2n)} dU \ \mbox{e}^{-\frac {V}{4}F_\pi^2\mu^2
{\rm Tr} [U,B][U^{-1},B]\ +\ \frac 12 \Sigma V {\rm Tr}M(U + U^{-1})}.
\label{zeff}
\ee
Note that the chemical potential and the quark masses only appear through the
dimensionless combinations
\be
\hatmu\equiv \mu F_\pi \sqrt{V} \ \ \ {\rm and} \ \ \ \hat M \equiv M
\Sigma V,  
\ee
where $V$ is the volume of space-time. 
In the absence of conjugate quarks 
the charge matrix $B$ is the unit matrix and
the fermionic partition function is independent of the chemical
potential. 
For example the partition function with one fermionic flavor is \cite{LS}
\be
Z^{N_f=1}(\hat{m}) = I_0(\hat{m}).
\ee

\begin{figure*}[ht]
  \unitlength1.0cm
    \epsfig{file=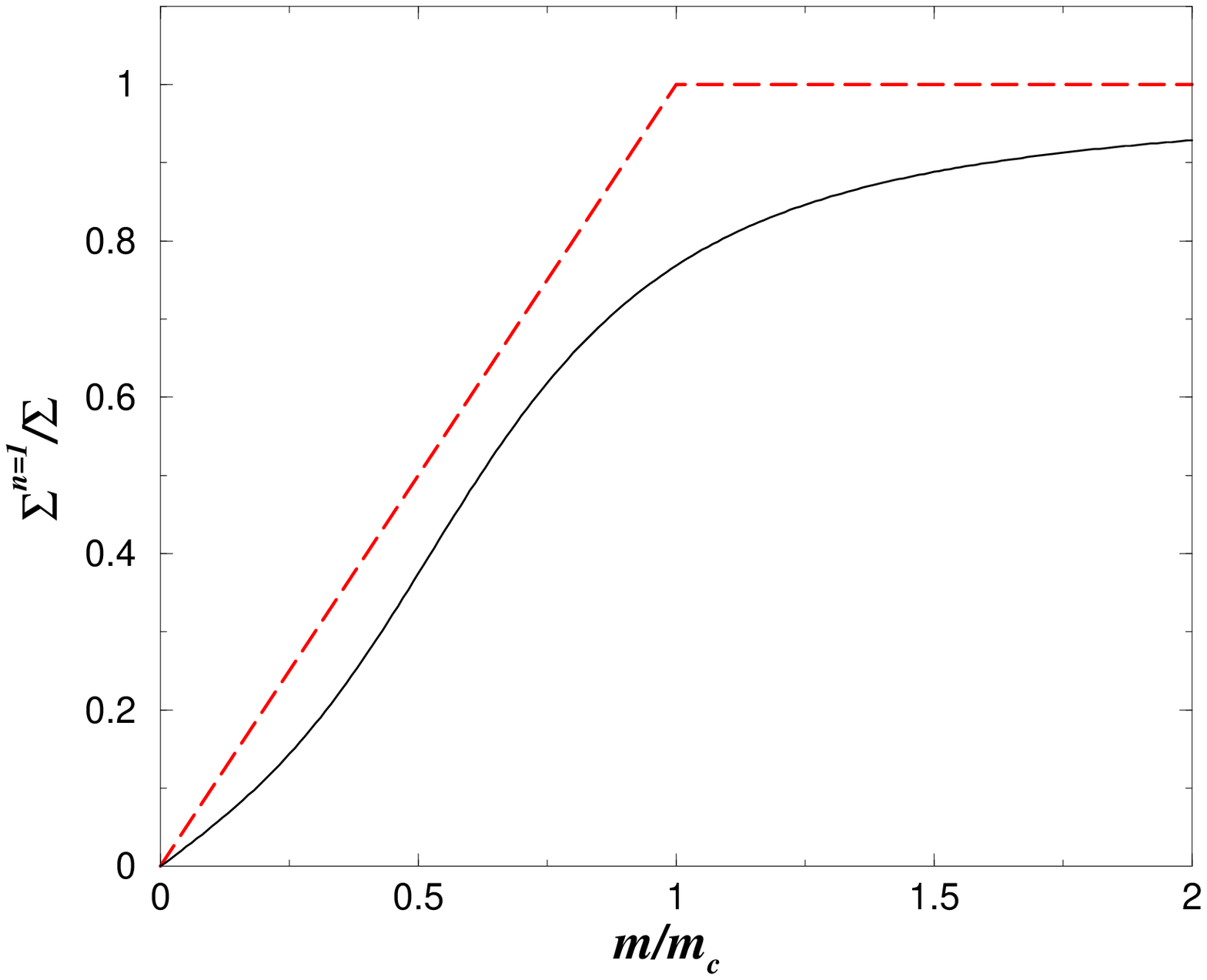,clip=,width=8.5cm}
    \epsfig{file=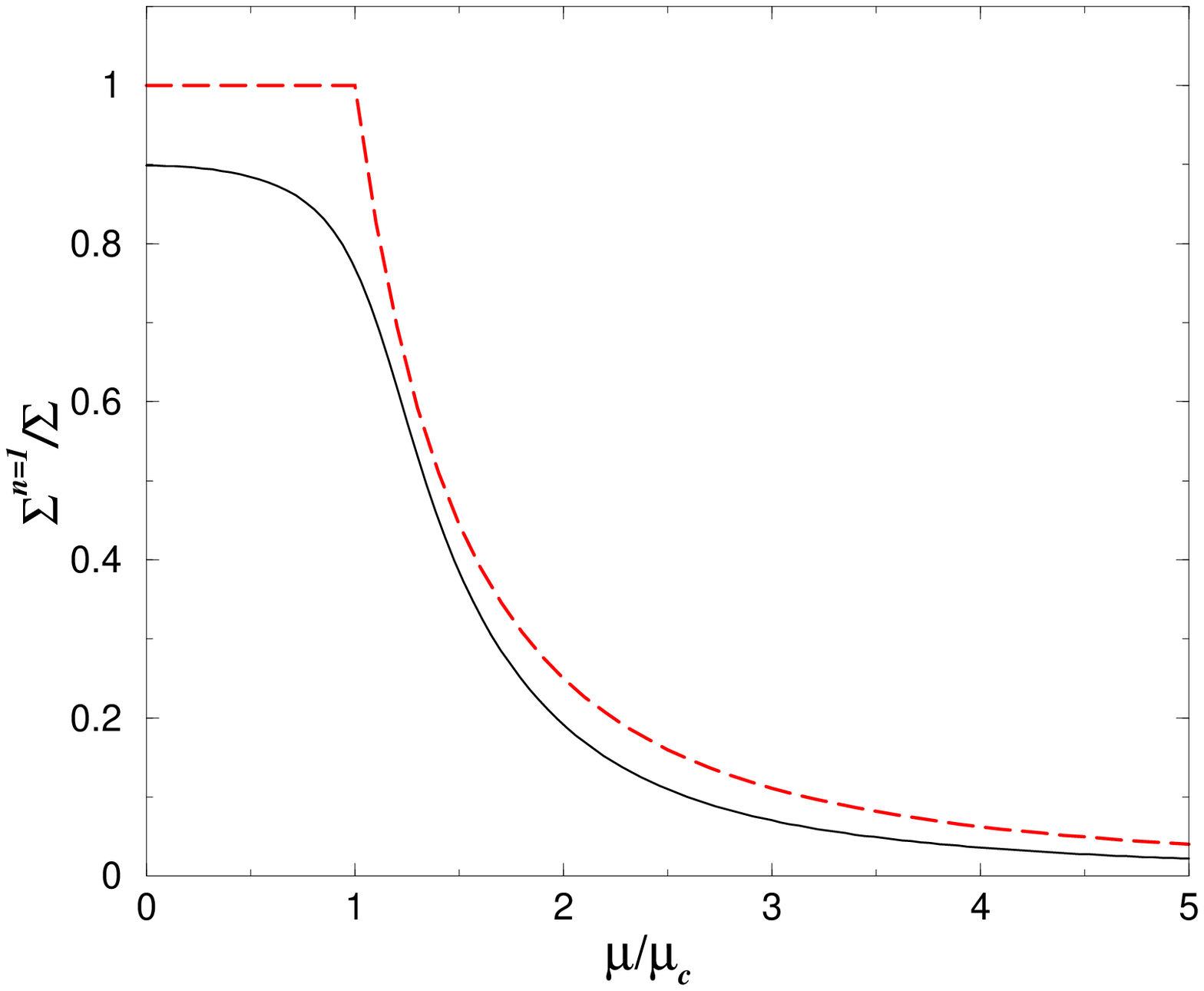,clip=,width=8.5cm}
  \caption{ 
  \label{fig:4}  The chiral condensate for $n=1$ (one fermion and one
  conjugate fermion). 
  {\bf Left:} The mass dependence for $\hat\mu=\sqrt{5}$ is shown by
  the full curve. In the thermodynamic limit ($\hat \mu \to \infty$, 
dashed curve) the
  condensate grows linearly with $m$ until it reaches a plateau at
  $m=m_c=2\mu^2F_\pi^2/\Sigma$.    
  {\bf Right:} The full curve displays the chemical potential
  dependence for $\hat{m}=10$ and the dashed curve shows the result in
  the thermodynamic limit ($\hat m \to \infty$).}
\end{figure*}

An explicit expression for  fermionic partition functions was derived in
\cite{SplitVerb2,AFV}. Here we take a closer look at the simplest nontrivial 
example which is given by the theory with one fermionic quark and one
conjugate fermionic quark both with  real mass $m$,    
\be
Z^{n=1}(\hat{m},\hat{m};\hatmu)  = 2\mbox{e}^{2\hatmu^2}
\int_0^1 dt \ t \ \mbox{e}^{-2\hatmu^2 t^2} I_0(\hat{m} t)^2.
\label{zn=1}
\ee
The chiral condensate follows by differentiation of the free energy,
\be
\frac{\Sigma^{n=1}(\hat{m};\hatmu)}{\Sigma} 
= \frac{1}{2}\del_{\hat{m}} \log Z^{n=1}(\hat{m},\hat{m};\hatmu). 
\ee
In figure \ref{fig:4} we have plotted the chiral condensate as a function of
$m$ (in units of $m_c=2\mu^2F^2/\Sigma$) and as a function of $\mu$
(in units of $\mu_c=\sqrt{m\Sigma}/(\sqrt{2}F_\pi)$). 
The full curves display a smooth dependence as was to be expected for a finite
volume. In the thermodynamic limit, the would-be phase transition
at $m = m_c$ in the left figure or at $ \mu =\mu_c$ in the
right figure becomes sharp. In terms of
the dimensionless variables this limit is reached for 
$\hat{m}\to\infty$ or $\hat \mu\to \infty$ and 
a kink develops  at the expected value of $\hat m /(2\hat\mu^2) = 1$.  

In order to see how the kink is recovered we take the thermodynamic limit of
(\ref{zn=1}) which is given by a leading order saddle point approximation.  
Using the asymptotic form
of the Bessel function we obtain 
\be
Z^{n=1}(\hat{m},\hat{m};\hatmu)  \sim \mbox{e}^{2\hatmu^2 }
\int_0^1 dt \ \frac{1}{\hat{m}} \ \mbox{e}^{-2\hatmu^2 t^2} e^{2t\hat{m}}.
\ee
A phase transition occurs when the saddle point, 
\be
t=\frac{\hat{m}}{2\hat \mu^2},
\ee
hits the integration boundary at $t =1$, 
that is when $\hat{m}=2\hatmu^2$. In the phase
where the saddle point is inside the boundary we obtain 
\be
\frac{\Sigma^{n=1}(\hat{m};\hatmu)}{\Sigma}\sim \frac{\hat{m}}{2\hatmu^2}. 
\ee  
If the saddle point is outside the integration domain the chiral 
condensate is simply given by
\be
\frac{\Sigma^{n=1}(\hat{m};\hatmu)}{\Sigma} = 1.
\ee
These results are shown by the dashed curves in figure \ref{fig:4} and 
are in agreement with the results obtained from chiral perturbation theory
\cite{KST,KSTVZ,TV,eff}.

\section{The theory with a pair of conjugate bosonic quarks}
 
We now turn to QCD with bosonic quarks at nonzero chemical potential and
start off with the theory with a pair of conjugate bosonic flavors
(\ref{zn=-1}). This theory is simpler than the theory with a single
bosonic quark since the latter involves both fermionic
and bosonic flavors (see (\ref{ZNf-1structure})).     

\vspace{2mm}

\subsection{Covariant Derivatives}

As discussed in section \ref{sec:gen}, the $n=-1$ partition function
diverges. To derive the effective partition function we therefore need to
take into account the convergence of the integrals in addition to the
symmetries. If we write the Dirac operator as
\be
D= \mat 0 & d \\ -d^\dagger & 0 \emat
\ee
the determinant in (\ref{zphq}) can be rearranged as
\be
\left | \begin{array}{cccc} 
        \epsilon  & 0 & z &d+\mu \\
        0 & \epsilon & -d^\dagger+\mu & z \\
     z^* & -d+\mu& \epsilon & 0 \\
   d^\dagger +\mu & z^* &0 &\epsilon  \end{array}
 \right |
=\left | \begin{array}{cccc}  
       \epsilon& z &d+\mu & 0\\
       z^*& \epsilon  &0 & d-\mu  \\
     d^\dagger+\mu&0& \epsilon & -z^* \\
   0&d^\dagger -\mu & -z & \epsilon 
\end{array} \right |.
\label{rear}
\ee 
For the purpose of studying the transformation properties, we rewrite
the determinant of the right hand side in a $2\times2 $ block notation
as
\be
\mat \zeta_1 &d+B_1
\\ d^\dagger +B_2 & \zeta_2 \emat.
\ee
This operator becomes locally gauge invariant under time dependent
but spatially constant flavor
gauge transformations
\be
\mat \zeta_1 &d+B_1
\\ d^\dagger +B_2 & \zeta_2 \emat
\to 
\mat v^{-1} & 0 \\0& u^{-1} \emat
\mat \zeta_1 &d+B_1
\\ d^\dagger +B_2 & \zeta_2 \emat 
\mat u & 0 \\ 0 & v \emat.
\label{transbos}
\ee
if $B_1$ and $B_2$ are transformed as
\be
B_1 \to v B_1 v^{-1} -  [\del_0 v]  v^{-1}\nn \\
B_2 \to u B_2 u^{-1}  + [\del_0 u] u^{-1}
\ee
and the mass matrices are transformed as
\be
\zeta_1 \to v \zeta_1 u^{-1}\nn\\
\zeta_2 \to  u \zeta_2 v^{-1}.
\ee
In \cite{SplitVerb2} we showed that the Goldstone manifold
is this case is given by $Gl(2)/U(2)$.  The transformation
(\ref{transbos}) induces the following transformation on
the Goldstone fields
\be
Q \to u Q v^{-1}.
\ee
Therefore,
\be
\del_0 Q \to \del_0 u 
Q v^{-1}+u \del_0 Q v^{-1}+ u Q \del_0 v^{-1},\nn\\
\del_0 Q^{-1} \to \del_0 v Q^{-1} u^{-1} 
+v \del_0 Q^{-1} u^{-1} +v Q^{-1} \del_0 u^{-1}.
\ee
One immediately sees that the covariant  combinations are
\be
\del_0 Q -Q B_1 - B_2 Q , \nn \\
\del_0 Q^{-1} + B_1 Q^{-1} + Q^{-1} B_2.
\ee
Of course $B_{1, \nu} = B_{2,\nu}  = \mu \delta_{\nu\, 0} \sigma_3\equiv\mu \delta_{\nu\, 0}B$.
We thus obtain the covariant derivatives
\be
\nabla_\nu Q = \del_\nu Q - \mu\delta_{\nu\, 0}\{Q,B\}, \qquad    
\nabla_\nu Q^{-1} = \del_\nu Q^{-1} +\mu\delta_{\nu\, 0} \{Q^{-1},B\}.
\label{cov-bos}
\ee

The chiral Lagrangian is the low-energy limit of QCD and should have
the same covariance properties as were derived above. Taking into
account terms to order $p^2$ in momentum counting we find the Lagrangian
\be
{\cal L} = -\frac {F^2}4 {\rm Tr} \nabla_\nu Q\nabla_\nu Q^{-1} -\frac i2
\Sigma {\rm Tr}M^T(Q - I Q^{-1} I)
\label{lag-bcb}
\ee
with 
\be
M = \mat \epsilon & z
\\ z^* & \epsilon \emat  \qquad {\rm and} 
\qquad I = \mat 0 & 1 \\ -1 & 0 \emat.
\label{M-bos}
\ee
As before, the pion decay constant is denoted by $F_\pi$ and the absolute
value of the 
chiral condensate is given by $\Sigma$.
We emphasize that the sign before the kinetic term is opposite to
the sign of the kinetic term 
in the fermionic Lagrangian. This sign enters to compensate
the sign due to the noncompactness of the modes in $Gl(2)/U(2)$ so that 
the 
kinetic terms pion fields have the correct sign \cite{foundations,zirn}.

In the case of the fermionic partition function with one flavor and
one conjugate flavor, we also could have rearranged the fermion determinant
as in (\ref{rear}), and we would have obtained covariant derivatives as
in (\ref{cov-bos}). Indeed, if we make the transformation 
\be 
U \to U\sigma_1   .
\label{UI}
\ee
in the Lagrangian (\ref{lferm}), the covariant derivatives in (\ref{covferm})
change to
\be
 \del_\nu U + \mu\delta_{\nu \, 0}[U,B] &\to& (\del_\nu U - \mu\delta_{\nu \, 0}\{U,B\})\sigma_1, \nn \\    
 \del_\nu U^{-1} +\mu\delta_{\nu \, 0}[U^{-1},B]&\to& \sigma_1(\del_\nu U^{-1} + \mu\delta_{\nu\, 0}\{U^{-1},B\} ).
\label{cov-fer}
\ee
Notice that both $U$ and  $U\sigma_1$ are unitary. 
The mass term in (\ref{lferm}) becomes
\be
{\rm Tr}\mat z & 0 \\ 0 &z^* \emat
(U +  U^{-1} ) ={\rm Tr}M(U +  \sigma_3 U^{-1} \sigma_3 ) \to 
{\rm Tr}\mat z & 0 \\ 0 &z^* \emat
 (U - \sigma_1 \sigma_3 U^{-1} \sigma_3 \sigma_1 )\sigma_1
={\rm Tr} \mat 0&z^*  \\ z&  0 \emat (U  +I U^{-1}I ).\nn \\
\ee
The invariance properties of this term are consistent with those of the
representation (\ref{rear}). Because both $U$ and $-U$ belong to 
$U(2)$ this term has the discrete symmetry $M \to -IMI$. In the
bosonic case this symmetry dictates that the relative sign of the two
terms in the mass term has to be negative.

\subsection{Microscopic limit of the partition function}

In the microscopic limit, the partition function is dominated by the
zero momentum modes and in the sector of zero topological charge it 
is given by
\be
Z^{n=-1}(z,z^*;\mu) = 
\lim_{\epsilon\to0} 
\int \frac{dQ}{{\det}^{2} Q} \theta(Q) \ 
\mbox{e}^{ -\frac{V}{4}F_\pi^2 \mu^2 {\rm Tr} \{Q ,B\}\{Q^{-1} ,B \}
\ +\ \frac{i}{2} {V\Sigma}{\rm Tr} M^T(Q -I Q^{-1}I )   } ,
\label{ZMINQCD}
\ee 
where $dQ\theta(Q)/{\det}^2 Q $ is the integration measure on positive
definite $2\times 2$ Hermitian matrices \footnote{In \cite{AOSV} we
incorrectly used commutators in (\ref{ZMINQCD}) instead of anti-commutators.
This resulted in an extra overall factor of $\exp(2V F^2_\pi \mu^2)$ which
was corrected for by hand in the calculation of the spectral density.
In \cite{SplitVerb2}, the correct form of $Z^{n=-1}(z,z^*;\mu) $ was
obtained because the overall $\mu$-dependent constant was  obtained 
by matching to the spectral density via the Toda lattice equation.}.
In the limit $\epsilon \to 0$ the integrals can be performed analytically
resulting in \cite{SplitVerb2}
\be \label{Znm1}
Z^{n=-1}(\hat{z},\hat{z}^*;\hat\mu)  = \log \epsilon \, 
\frac{e^{-2{\hat{\mu}}^2}}{\hat\mu^2}
\exp\left(- \frac{\hat{z}^2 + \hat{z}^{*\,2}}{8\hat\mu^2}  \right)
K_0 \left( \frac{|\hat{z}|^2}{4 \hat\mu^2} \right ).  
\ee
Using the Toda lattice equation, the product of this partition
function and its fermionic counterpart gives the quenched microscopic
spectral density at nonzero chemical potential. This has been
confirmed beautifully by recent quenched lattice QCD simulations 
at nonzero chemical potential with 
a staggered Dirac operator \cite{tilo-james} and an overlap Dirac operator
\cite{tilo}.

\begin{figure*}[ht]
  \unitlength1.0cm
    \epsfig{file=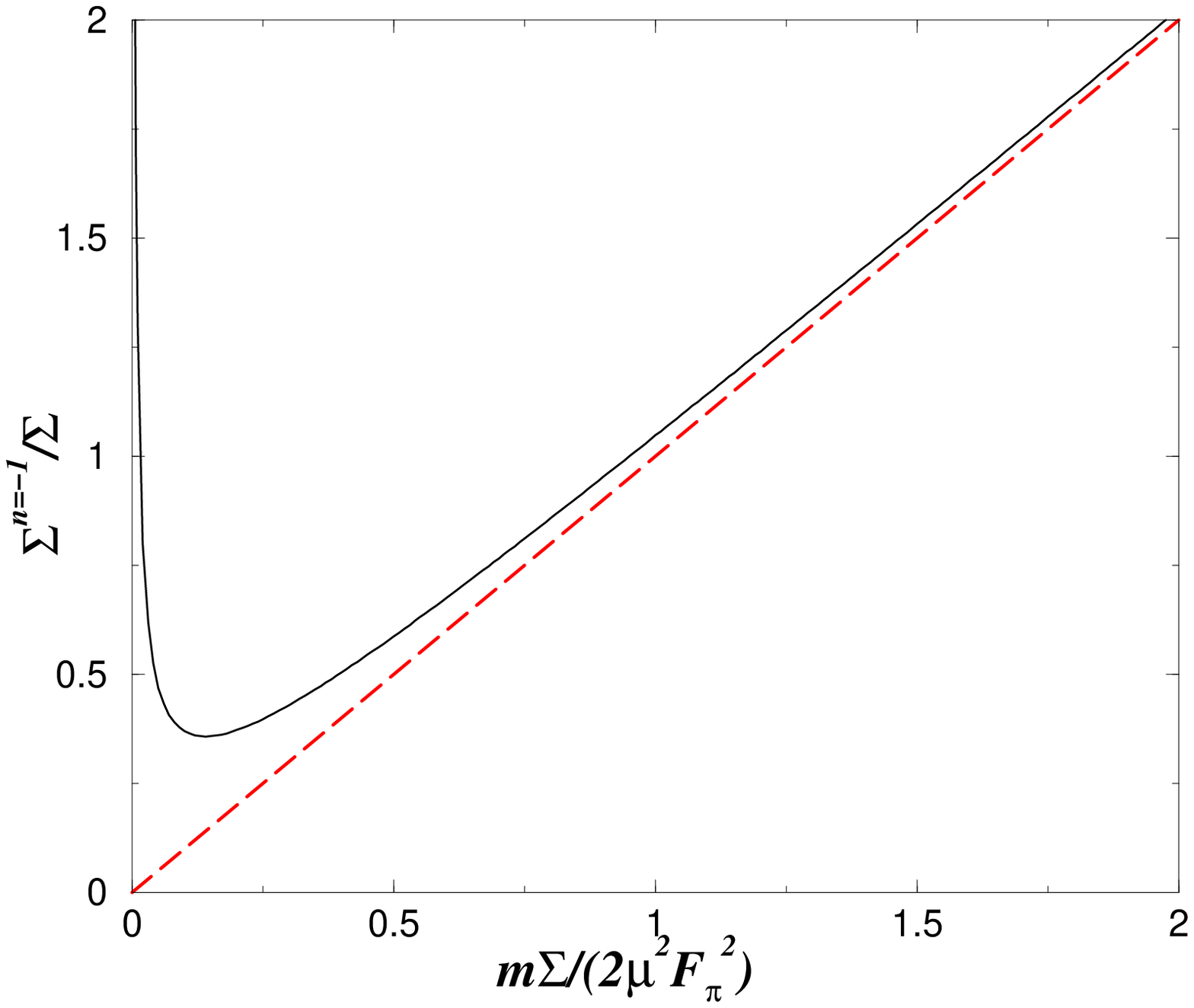,clip=,width=8.5cm}
    \epsfig{file=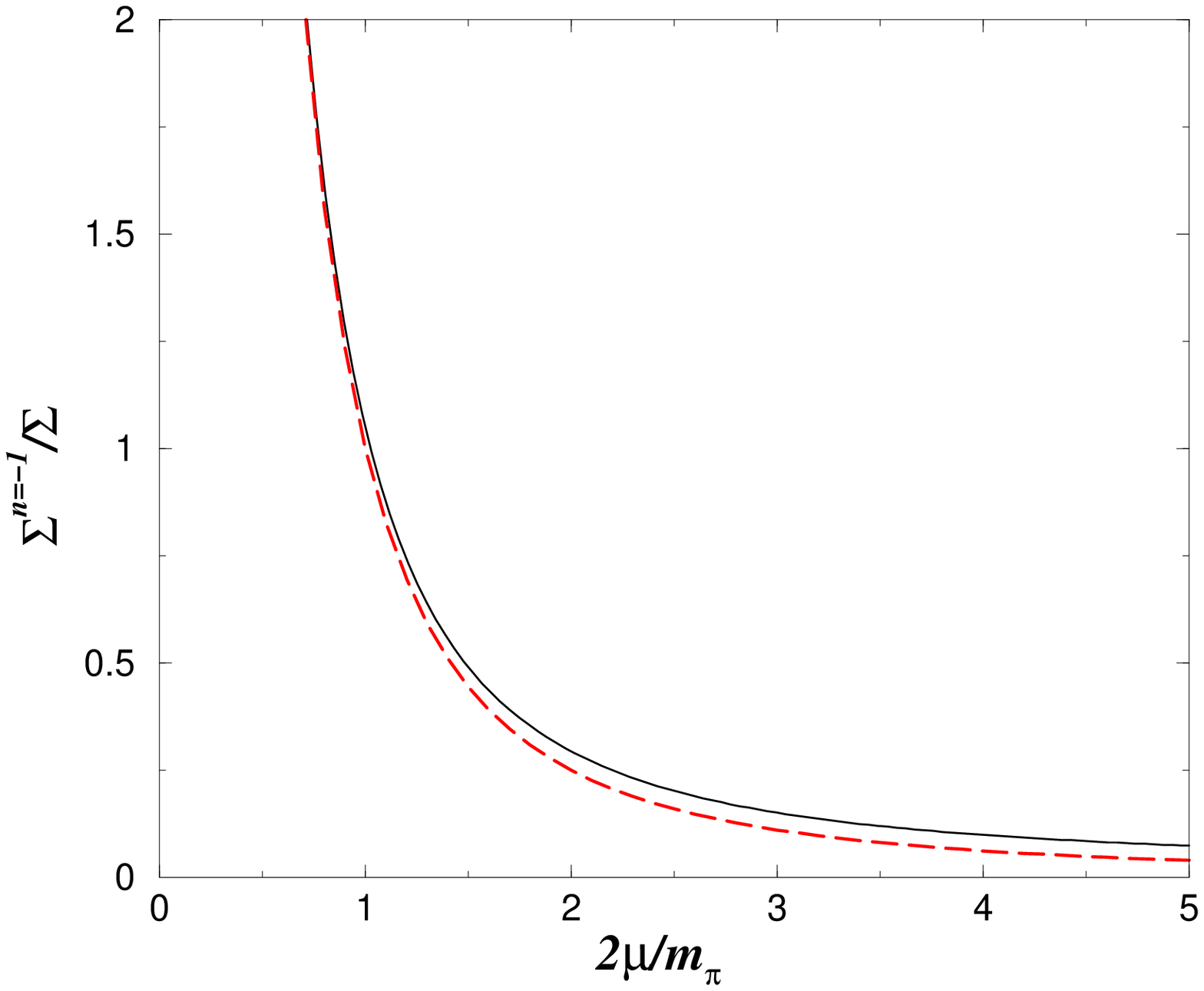,clip=,width=8.5cm}
  \caption{ 
  \label{fig:2}  The chiral condensate for $n=-1$ (one boson and one
  conjugate boson). 
  {\bf Left:} The mass dependence for $\hat\mu=\sqrt{5}$ (full curve).  
  {\bf Right:} The chemical potential dependence for $\hat{m}=10$
  (full curve). 
  In this case there is no phase transition  at $\mu=m_\pi/2$ as can
  be seen from the result in the thermodynamic limit (dashed curves,
obtained for $\hat \mu \to \infty $ or $\hat m \to \infty$, respectively).} 
\end{figure*}

The volume dependence has dropped from the partition function 
(\ref{Znm1}) and
no kink will develop in the thermodynamic limit. 
The chiral condensate is given by
\be
\frac{\Sigma^{n=-1}(\hat{m};\hat\mu)}{\Sigma} 
&=& -\frac{1}{2}\del_{\hat{m}} \log Z^{n=-1}(\hat{m},\hat{m};\hat\mu) 
\nn \\ &\sim&
\frac{\hat{m}}{2\hat\mu^2}\qquad {\rm for } \qquad 
\hat{m}\to\infty \ \ , \ \ \hat\mu \to \infty \ \ {\rm and}
 \ \  \frac{\hat m}{2\hat\mu^2} \sim 1. 
\ee
It follows that the low energy theory with one 
bosonic flavor and its conjugate does {\it not} have a phase transition 
as function of $\hat\mu$. 
This is to be contrasted with
the fermionic counterpart which, as discussed in section \ref{sec:ferm}, has
a phase transition  at $\hat{m}=2\hat\mu^2$. Plots of the chiral condensate
for $n=-1$ are shown in Fig.~\ref{fig:2}.

The reason for the absence of a phase transition 
at $\hat m =2\hat\mu^2$ 
becomes clear from
the calculation of the mass spectrum of the Lagrangian (\ref{lag-bcb})
along the lines of \cite{KSTVZ} (see Appendix A). 
For all values of the chemical potential
we find a charged massless mode (in fact with a mass $\sim \sqrt \epsilon$) 
which condenses for any $\mu >0$. The masses of the remaining three modes
are given by $2\mu$, $m_\pi^2/2\mu$ and $2\mu \sqrt{1 +3(m_\pi/2\mu)^4}$,
exactly the same as in other QCD-like theories at nonzero chemical 
potential \cite{KSTVZ,eff}.

\vspace{2mm}

\section{The effective partition function and chiral random matrix theory}

We now turn to the computation of the partition function with one bosonic
quark, that is $N_f=-1$. As follows from eq.~(\ref{ZNf-1structure}) the 
Goldstone bosons result from the breaking of chiral symmetry in the 
theory with one fermionic and two bosonic
quarks.  The partition function is therefore given by an integral over the
super group $\hat{Gl}(1|2)$. Such integrals are technically rather
complicated, and
at this point we have not succeeded to evaluate the
 $N_f=-1$ partition function along these lines.

Fortunately, the effective low energy partition function has an
alternative representation as the large $N$ limit of a
random matrix theory with the same chiral symmetries.
This has been shown explicitly 
in the fermionic case at zero \cite{SV} and nonzero
\cite{SplitVerb2,O,AFV} chemical  potential as well as for bosonic \cite{DV} 
and supersymmetric \cite{OTV,DOTV,AF2,FA} partition functions at zero chemical
potential. For bosonic quarks at nonzero chemical potential the equivalence
between the effective theories and random matrix theory in the microscopic
limit has been established for $n=-1$ in \cite{AOSV}.   
 
As before, we only consider the theory in the sector of zero
topological charge.

\section{The Random Matrix Model}
\label{sec:rmt}

The random matrix partition function with $N_f$ quark flavors of mass $m$
and $n$ pairs of regular and conjugate quarks with masses $y$ and
$z^*$, respectively, is defined by \cite{O} 
\be
{\cal Z}_N^{N_f,n}(\{m_f\},y,z^*;\mu) &\equiv&
 \int d\Phi  d\Psi \ w_G(\Phi)  w_G(\Psi) 
{\det}^{N_f}(\,{\cal D}(\mu) + m_f\,) \  \nn \\
&&\times
{\det}^n(\,{\cal D}(\mu) + y\,){\det}^n(\,{\cal D}^\dagger(\mu) + z^* ) ,
\label{ZNfNb}
\ee
where the non-Hermitian Dirac operator is given by
\be
\label{dnew}
\mathcal{D}(\mu) = \left( \begin{array}{cc}
0 & i \Phi + \mu \Psi \\
i \Phi^{\dagger} + \mu \Psi^{\dagger} & 0
\end{array} \right) ~.
\ee
Here, $\Phi$ and $\Psi$ are complex $N\times N$  matrices with the same
Gaussian weight function 
\be
\label{wg}
w_G(X) ~=~ \exp( \, - \, N \, \tr \, X^{\dagger} X \, ) ~.
\ee 
Bosonic quarks appear as inverse determinants and 
notationally we simply allow $N_f$ and $n$ to take negative values. 

\vspace{2mm}

Of course, in general the QCD partition function and the random matrix
partition function are different. 
However, 
when we consider the microscopic limit where the variables
\be
\hat{m} = 2 m N   \ \ \ {\rm and} \ \ \ \hat{\mu}^2=2\mu^2N 
\ee    
are fixed as $N\to\infty$ the random matrix partition function and the QCD
partition function coincide provided that we identify (see the discussion 
in \cite{AOSV}) 
\be\label{mapping}
\hat{m}= 2 m N & \to &  m V \Sigma \\
\hat{\mu} = 2 \mu^2 N & \to &  \mu^2 F_\pi^2 V. \nn
\ee
In this section we will work within the random matrix framework
and use the identifications (\ref{mapping}) in the final results.
 
\vspace{2mm}

In \cite{O} it was shown that the random matrix partition function
(\ref{ZNfNb}) can be rewritten in the eigenvalue representation,
\be
\label{epfnew}
{\cal Z}_N^{N_f,n}(m,y,z^*;\mu)  \sim
 \int_{\mathbf{C}} \prod_{k=1}^{N} d^2z_k \,
{\cal P}^{N_f,n}(\{z_i\},\{z_i^*\}; \mu),
\ee
where the  integration extends over the full complex plane and 
the joint probability distribution of the eigenvalues is given by
\be
\label{jpd}
{\cal P}^{N_f,n}(\{z_i\},\{z_i^*\};\mu)
&=& \frac{1}{\mu^{2N}}\left|\Delta_N(\{z_l^2\})\right|^2 \, 
\prod_{k=1}^{N} w(z_k,z_k^*;\mu) (m^2-z^2_k )^{N_f} (y^2-z_k^2 )^{n}
(z^{*\,2}-z_k^{*\,2} )^{n} .
\ee 
The Vandermonde determinant is defined as  
\be
\Delta_N(\{z^2_l\}) \equiv \prod_{i>j=1}^N (z_i^2-z_j^2),
\label{vander}
\ee
and the weight function reads
\be 
w(z_k,z^*_k;\mu) &=& |z_k|^{2} 
K_0 \left( \frac{N (1+\mu^2)}{2 \mu^2} |z_k|^2 \right)
\exp\left(-\frac{N (1-\mu^2)}{4 \mu^2}  
(z^2_k + \conj{z_k}^2) \right). 
\label{wnew}
\ee
The eigenvalue representation makes it possible to employ the method of 
orthogonal polynomials in the complex plane \cite{A03,AV,BI,BII,AP} to compute 
the spectral density and  eigenvalue correlation functions \cite{O}.

\subsection{Orthogonal Polynomials and their Cauchy transform}

In order to evaluate the partition function with $N_f=-1$ we will make
use of orthogonal polynomials and their Cauchy transform. The complex
Laguerre polynomials given by \cite{O}
\be
p_k(z;\mu) = \left( \frac{1-\mu^2}N\right )^k k! 
L_k \left ( -\frac{Nz^2}{1-\mu^2} \right)
\ee 
are the orthogonal polynomials corresponding to the weight $w(z,z^*;\mu)$
given in (\ref{wnew}). 
To be specific, the polynomials satisfy the orthogonality condition 
\cite{Akemann} 
\be
\int_{\mathbf{C}}d^2z\ w(z,z^*;\mu)\ p_k(z;\mu)\ p_l(z;\mu)^* ~ 
 \ =\ r_k ~ \delta_{kl} \ ,
\label{OPdef}
\ee
with the norm 
\be
\label{Norm}
r_k ~=~
\frac{  \pi \, \mu^2 ~ (1+\mu^2)^{2k} ~ k! ~ k!}
     { N^{2k +  2}}  ~.
\ee
The Cauchy transform of the orthogonal polynomials is defined as
\be
h_k(m;\mu) = \int_{\bf C} d^2z \frac 1{z^2-m^2}w(z,z^*;\mu) p^*_k(z;\mu),
\ee
where ${\bf C}$ indicates that the integration extends over the complex
plane. Using that the weight function and polynomials are even functions of 
$z$ the Cauchy transform can be written as
\be 
h_k(m;\mu) = \int_{\bf C} d^2z \frac{1}{z(z-m)}w(z,z^*;\mu) p^*_k(z;\mu) .
\ee

\section {QCD with one bosonic flavor}

It was shown in \cite{AOSV} that
\be\label{ZNf-1:general}
Z_N^{N_f =-1}(m;\mu) = -\frac 1{r_{N-1}} h_{N-1}(m;\mu).
\ee
Therefore, studying the properties of bosonic partition
functions is equivalent to analyzing the properties of the 
Cauchy transform. While the relation between the partition 
function and the Cauchy transform was established in \cite{AOSV}, 
no explicit evaluation of the Cauchy transform was given; this evaluation
follows below. We will find that the partition function with one bosonic quark
depends on the chemical potential for $\mu > m_\pi/2$.

\vspace{2mm}

We are interested in the microscopic limit where $N\to \infty$ for
fixed $\hat{m} = 2Nm$ and fixed $\hat{\mu}^2 = 2\mu^2N$. (This scaling of
$\mu$ is the analogue of the weak non-hermiticity limit introduced in 
\cite{FKS}.) 
In this limit the polynomial $p_{N-1}/r_{N-1}$ is given by
\be 
\frac{p_{N-1}(\hat{z};\hatmu)}{r_{N-1}} &=& \frac 1{r_{N-1}} 
\frac {(N-1)!}{N^{N-1}}(1-\frac{\hat\mu^2}{2N})^{N-1} 
L_{N-1}\left( -\frac{\hat{z}^2}{4N}\right )\nn \\
&\sim & \frac {2N^{5/2}e^N} {\sqrt{2\pi}\pi \hat\mu^2} 
e^{-\frac 32 \hat\mu^2} I_0(\hat{z}),
\ee
where we have used that
\be
r_{N-1} = \frac{\pi\hat\mu^2(1+\hat\mu^2/2N)^{2(N-1)} (N-1)!(N-1)!}
{2N^{2N+1}}.
\ee 
The microscopic limit of the partition function is thus given by
\be
Z^{N_f=-1}(\hat m;\hat\mu) = -\frac {2N^{1/2}e^N} {\pi {\hat\mu}^2} 
e^{-\frac 32 \hat\mu^2}
\int_{\bf C} d^2z \frac {z^*}{z-\hat{m}} 
K_0\left (\frac{|z|^2}{4\hat\mu^2}\right )
e^{-\frac{z^2+z^{*\, 2} } {8\hat\mu^2}} I_0(z^*).
\label{znfm1}
\ee
To calculate this integral we write (with $z=x+iy$)
\be
I_0(z^*) = {\rm Sgn}(y)\frac 1{\pi i} (K_0(z^*) - K_0(-z^*))
\label{ikk}
\ee
and first calculate the integral over $y$ by a complex
contour integration. The first term in (\ref{ikk}) is exponentially
damped in the upper half of the complex $y$-plane 
and the second term in the
lower half-plane. This allows us to close the integration contour of the $y$
variable, and by Cauchy's theorem we obtain for the real part of the partition
function 
\be
{\rm Re }(Z^{N_f=-1}(\hat{m};\hat\mu))= && 2\pi c_{N}(\hat\mu) 
\int_{-\infty}^\infty dx (2x-\hat{m})
\exp[-\frac{2x^2-2x\hat{m}+\hat{m}^2}{4\hat\mu^2}]  
\nn \\  \times &&
\left [ 
\theta (\hat{m}-x)
\theta(2x-\hat{m})K_0\left ( \frac{(2x-\hat{m})\hat{m}}{4\hat\mu^2}\right )
I_0(2x-\hat{m})\right . 
\nn\\
&& \left . - \theta(\hat{m}-x)\theta(\hat{m}-2x) I_0\left 
( \frac{(2x-\hat{m})\hat{m}}{4\hat\mu^2}\right )K_0(\hat{m}-2x)
\right ].
\ee 
The imaginary part of the partition function is given by
\be 
{\rm Im} (Z^{N_f=-1}(\hat{m};\hat\mu))= &&-      
2 c_{N}(\hat\mu) \int dx (2x-\hat{m})
\exp[-\frac{2x^2-2x\hat{m}+\hat{m}^2}{4\hat\mu^2}]  
\nn \\ \times  &&
\left [ \theta(x-\hat{m})
 K_0\left ( \frac{(2x-\hat{m})\hat{m}}{4\hat\mu^2}\right )K_0(2x-\hat{m})\right .
\nn \\   
&&+\theta (\hat{m}-x)
\theta(2x-\hat{m})K_0\left ( \frac{(2x-\hat{m})\hat{m}}{4\hat\mu^2}\right ) K_0(2x-\hat{m})\nn\\
&& \left . + \theta(\hat{m}-x)\theta(\hat{m}-2x) K_0\left 
( \frac{(\hat{m}-2x)\hat{m}}{4\hat\mu^2}\right )K_0(\hat{m}-2x)
\right ].
\ee 
The overall constant $c_N(\hat \mu)$ is defined by
\be
c_N(\hat \mu)=\frac {2N^{1/2}e^N} {\pi {\hat\mu}^2} 
e^{-\frac 32 \hat\mu^2}.
\label{cnhat}
\ee
The integrand of the imaginary part is odd about $x=\hat{m}/2$ so that
the integral vanishes.
Using $s = x-\hat{m}/2$ as new integration variable we obtain the final
expression 
for the partition function with $N_f=-1$ at nonzero chemical potential
\be
Z^{N_f =-1}(\hat{m};\hat\mu) = && 4\pi c_N(\hat\mu) e^{-\frac{\hat{m}^2}{8\hat\mu^2}}
\int_{-\infty}^{\hat{m}/2} ds s \exp[ -\frac {s^2}{2\hat\mu^2}]
\nn \\   \times
&& \left [
\theta(s)K_0\left ( \frac{s\hat{m}}{2\hat\mu^2}\right ) I_0(2s) -
\theta(-s) I_0\left 
( \frac{s\hat{m}}{2\hat\mu^2}\right )K_0(-2s)
\right ]\nn \\
= 
&& 4\pi c_N(\hat\mu) e^{-\frac{\hat{m}^2}{8\hat\mu^2}}\nn \\ \times && \left [
\int_{0}^{\hat{m}/2} ds s \exp[ -\frac {s^2}{2\hat\mu^2}]
K_0\left ( \frac{s\hat{m}}{2\hat\mu^2}\right ) I_0(2s) 
+\int_{0}^{\infty} ds s \exp[ -\frac {s^2}{2\hat\mu^2}]
I_0\left 
( \frac{s\hat{m}}{2\hat\mu^2}\right )K_0(2s) \right ].
\label{s-integral}
\ee

\subsection{The chiral condensate for one bosonic flavor}

\begin{figure}[ht]
  \unitlength1.0cm
    \epsfig{file=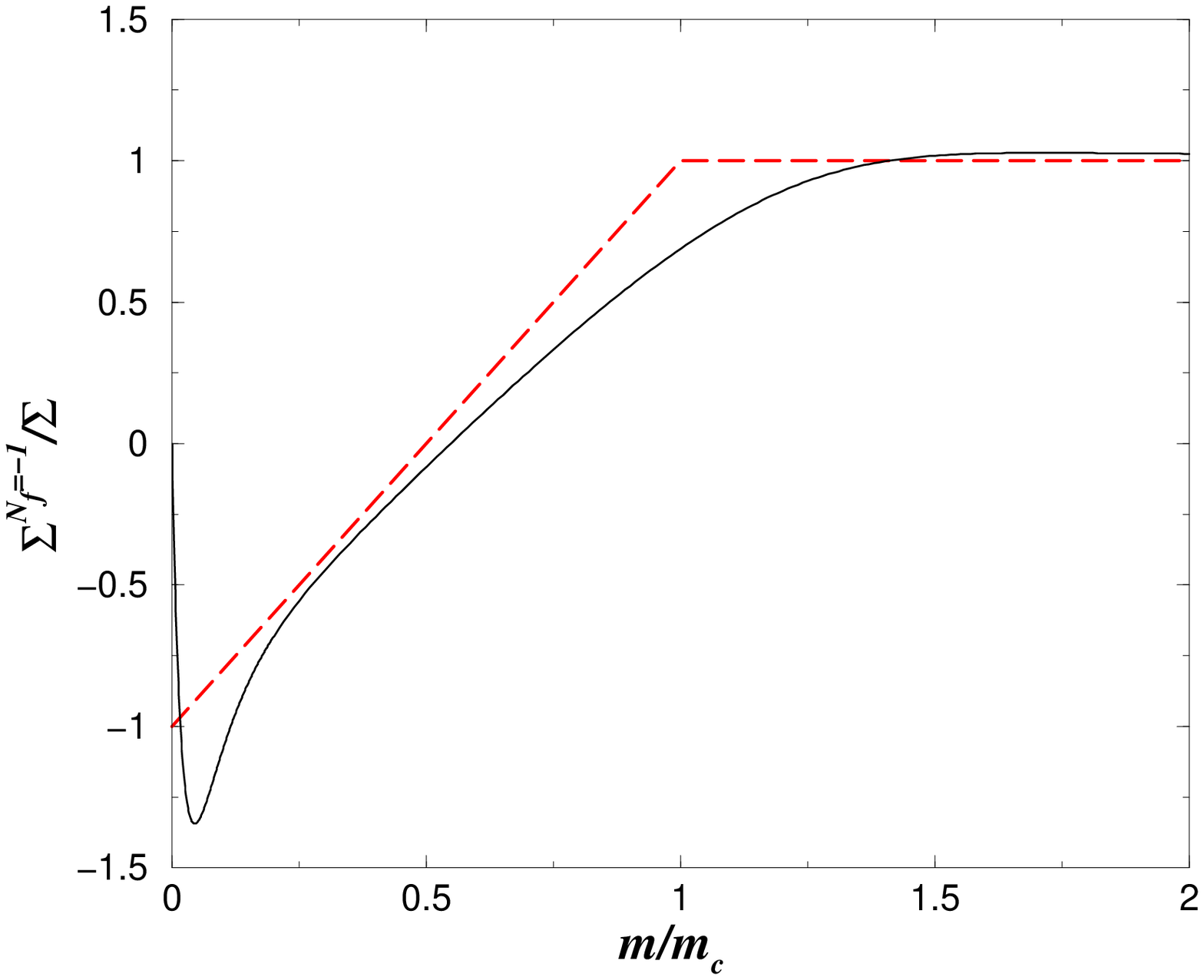,clip=,width=8.5cm}
    \epsfig{file=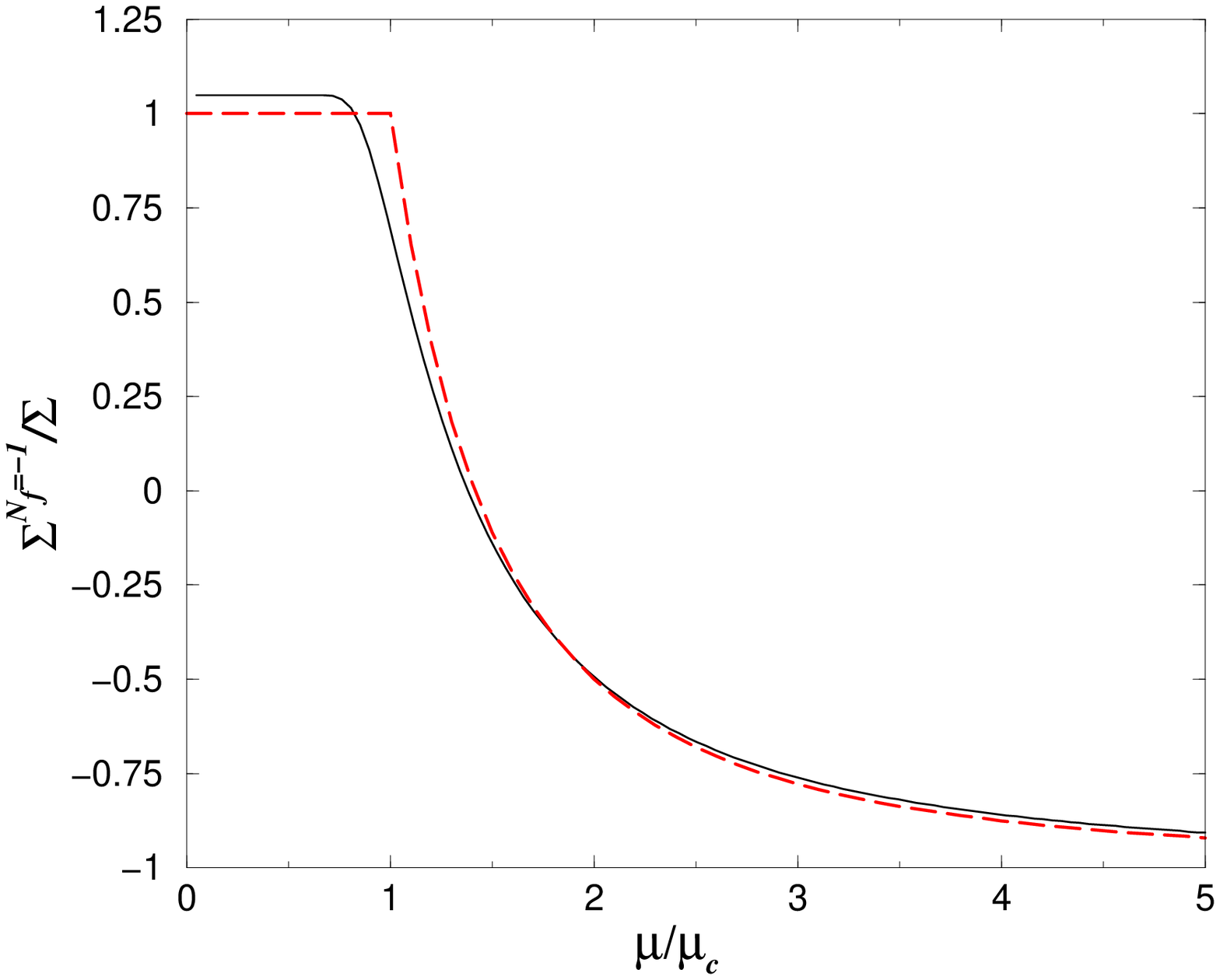,clip=,width=8.5cm}
  \caption{ 
  \label{fig:1} The chiral condensate in the theory with one bosonic flavor. 
   {\bf Left:} the mass dependence for $\hat\mu=\sqrt{5}$ (full
   curve). In the thermodynamic limit ($\hat \mu \to \infty$, dashed curve) 
$\Sigma^{N_f=-1}$
   changes from a linear  dependence on $m$ to become $m$ independent for
   $m>m_c=2\mu^2F_\pi^2/\Sigma$.
   {\bf Right:}  The dependence on the chemical potential for
   $\hat{m}=10$ (full curve) and in the thermodynamic limit 
($\hat m \to \infty$,  dashed curve).}
\end{figure}

The chiral condensate given by
\be
\frac{\Sigma^{N_f=-1}(\hat{m};\hat\mu)}{\Sigma} 
= -\d_{\hat{m}} \log Z^{N_f=-1}(\hat{m};\hat\mu)
\label{SigmaNf=-1def1}
\ee
is plotted in figure \ref{fig:1} (full curve). Notice that the chiral
condensate depends on the chemical potential, and that for
$\hat{m}\gg2\hat\mu^2$, it approaches the value $\Sigma$. Below we
will show that in the thermodynamic limit $\Sigma^{N_f=-1}$ develops a
kink at $\hat{m}=2\hat\mu^2$.   

\subsection{Phase transition with one bosonic flavor}

Above we have 
determined the partition function in the microscopic limit, where the
source $m$ times the volume $N$, is kept fixed, and as a consequence, 
the phase transition
at $\hat{m}=2\hat\mu^2$ is smeared. 
A cusp in the derivative of the free energy
only appears in the thermodynamic limit where $N\to\infty$ for fixed
$m$.  This limit can also be approached by  taking
$\hat{m}=2mN\to\infty$ and $\hat\mu^2=2\mu^2 N\to\infty$ 
in the microscopic results as will be done below.  

\vspace{2mm}

In the thermodynamic limit the integral (\ref{s-integral}) can be 
calculated by a saddle point approximation.
The saddle point of the first integral is at
\be\label{saddle1}
 s =  -\frac{\hat{m}}{2} +2\hat\mu^2,
\ee
and the saddle point of the second integral is at
\be
s = -2\hat\mu^2 +\frac{\hat{m}}{2}.
\ee
In the thermodynamic limit the partition function 
for $\hat m > 2{\hat \mu}^2$ is therefore  given by
\be
Z^{N_f=-1}(\hat{m};\hat\mu) =\frac{4\pi}{\sqrt{2\hat{m}}} c_N(\hat \mu) \mu^2  e^{2\hat{\mu}^2-\hat{m}}.
\ee
The exponent
depends linearly on $\hat{m}$ resulting in a chiral condensate that 
is equal to $1$. In the normalization for which the partition function
is $\mu$-independent 
for small $\mu$ (see \cite{O}),  the exponent in $c_N(\hat \mu)  $
(see eq.~(\ref{cnhat}))
becomes $ \exp (-2\hat \mu^2) $ instead of $ \exp(-\frac 32 \hat\mu^2)$.
A phase transition occurs when the dominant saddle point hits the
boundary of the integration domain, i.e.~at $\hat{m}=2\hat\mu^2$,
cf.~(\ref{saddle1}). 
For $\hat{m}<2\hat\mu^2$ the saddle point is outside of the domain of the
first integral and the main contribution to the first integral
comes from the region close to $\hat{m}/2$ so that
\be
Z^{N_f=-1}(\hat{m};\hat\mu) \sim e^{\hat{m} -\hat{m}^2/2\hat\mu^2-2\hat\mu^2}
\qquad {\rm for } \quad \hat\mu^2 > \frac{\hat{m}}{2}.
\ee
The condensate in the thermodynamic limit is therefore given by
\be
\frac{\Sigma^{N_f=-1}(\hat{m};\hat\mu)}{\Sigma} = \left\{
\begin{array}{ccc}
-1+\frac{\hat{m}}{\hat\mu^2} & {\rm for} & \hat{m}<2\hat\mu^2 \\
1 & {\rm for} & \hat{m}>2\hat\mu^2 
\end{array}\right..
\ee
In figure~\ref{fig:1}, this result is shown by the dashed curves.

As promised, we have shown that in the thermodynamic limit
the chiral condensate has a kink at $\hat{m}=2\hat\mu^2$.
Using (\ref{mapping}) and the GOR-relation
($2m\Sigma=F_\pi^2m_\pi^2$) we see 
that the phase transition takes place at $\mu=m_\pi/2$ with
\be
\frac{\Sigma^{N_f=-1}(m_\pi;\mu)}{\Sigma} = \left\{
\begin{array}{ccc}
-1+\frac{m_\pi^2}{2\mu^2} & {\rm for} & m_\pi < 2\mu \\
1 & {\rm for} &  m_\pi > 2\mu 
\end{array}\right..
\ee
This behavior of the chiral condensate can be explained from the
structure in (\ref{ZNf-1structure}). We first note that the average mass 
derivatives
of fermionic and bosonic determinants give factors that are equal in
magnitude but have opposite sign.
At $\mu = 0$ the contribution from the fermionic quark thus cancels against
the contribution 
from one of the bosonic quarks resulting in a total chiral condensate of
magnitude $\Sigma$. For $\mu<m_\pi/2$ the partition function is $\mu$
independent and the chiral condensate therefore remains at the value
$\Sigma$. As $\mu$ exceeds $m_\pi/2$ it becomes energetically favorable to
create pions made out a bosonic anti-quark and a conjugate bosonic quark.    
As a consequence the contribution to the chiral condensate from the 
bosonic quarks starts rotating to zero for $\mu>m_\pi/2$ so that the 
total chiral condensate for large values of $\mu/m_\pi$ is given by 
the contribution of the fermionic flavor alone. For $\mu\gg m_\pi/2$ 
the chiral condensate therefore is equal to the chiral condensate at 
$\mu =0$ but with the opposite sign.

The second order phase transition can also be seen in the baryon density given
by 
\be
\frac{n^{N_f=-1}(\hat{m};\hat\mu)}{F_\pi} = -\del_{\hat\mu}\log Z^{N_f=-1}(\hat{m};\hat\mu) &=& \left\{
\begin{array}{ccc} 
0 & {\rm for} &  \hat\mu^2 < \frac{\hat{m}}{2} \\
4\hat\mu(1 - \frac {\hat{m}^2}{4\hat\mu^4}) & {\rm for} &  \hat\mu^2 > \frac{\hat{m}}{2}
\end{array}\right. .
\ee
In contrast, the free energy for $N_f=1$ does not depend on $\hat\mu$.

When $\hat{m}\sim 1$ the finite volume effects for the chiral condensate are
strong and the microscopic prediction differs significantly from the
result in the thermodynamic limit. For $\hat{m}\ll1$ the partition function
is given by 
\be
Z^{N_f=-1}(\hat{m};\hat\mu) = A + \frac{\hat{m}^2}8 \log \hat{m} + O(\hat{m}^2)
\ee
resulting in a chiral condensate
\be
\frac{\Sigma^{N_f=-1}(\hat{m};\hat\mu)}{\Sigma} = \frac 1{4A}  {\hat{m}\log \hat{m}} 
\ \ \ {\rm for} \ \ \ \hat{m}\ll1,
\ee
where $A$ is a constant. This explains the approach to zero for $\hat{m}\to0$
of the solid curve in the left panel of Fig.~\ref{fig:1}.

\section{The phase structure and the eigenvalue spectrum}

The chiral phases of QCD reflect themselves in the eigenvalue spectrum of
the Dirac operator. For zero chemical potential the eigenvalue density near
the origin of the imaginary axis is an order parameter for spontaneous
breaking of chiral symmetry \cite{BC}. When the chemical potential is nonzero
the eigenvalues move into the complex plane and the chiral condensate 
in full QCD is
then linked to the oscillations of the eigenvalue density \cite{OSV}. 
For nonzero isospin chemical potential ($n=1$) the transition into the
Bose-Einstein condensed phase takes place when the quark mass is at the
boundary of the support of the eigenvalue density (see for example 
\cite{TV}).

Also in the case with one pair of conjugate bosons, $n=-1$,
the phase structure is closely related to the Dirac spectrum. 
Setting $n=-1$ in (\ref{epfnew}) we obtain the following 
eigenvalue representation
of the partition function,
\be
\label{znm1}
{\cal Z}_N^{N_f=0,n=-1}(z,z^*;\mu)  \sim
\frac{1}{\mu^{2N}} \int_{\mathbf{C}} \prod_{k=1}^{N} d^2z_k 
\left|\Delta_N(\{z_l^2\})\right|^2 \, 
\prod_{k=1}^{N} w(z_k,z_k^*;\mu)\frac 1{ (z^2-z_k^2 )
(z^{*\,2}-z_k^{*\,2} )} .
\ee 
We observe that the probability to find an eigenvalue, 
$z_j$, has a modulus squared pole 
at the mass $z$. This gives rise to the logarithmic divergence discussed in
section II,
\be
\int_{C_\epsilon(u)} d^2 u \frac 1{|z^2-u^2|^2} \sim \frac 1{|z|^2} \log \epsilon.
\label{ediv}
\ee
Here, $C_\epsilon(u)$ is an annulus centered at $u$ with outer radius 1
and inner radius $\epsilon \ll 1$.
Because of the Vandermonde determinant, the probability
that two or more eigenvalues are close to the mass $z$ does not diverge. 
Therefore the logarithmically divergent contribution to the partition function
is from eigenvalue configurations with exactly {\it one} eigenvalue close
to the mass.  
Setting $z_N=z$ in (\ref{znm1}) and using that 
\be
\left.|\Delta(\{z_l^2\}_{l=1}^N)|^2\right|_{z_N=z} =
|\Delta(\{z_l^2\}_{l=1}^{N-1})|^2 \prod_{l=1}^{N-1}|z_l^2-z^2|^2, 
\ee
we find that the product on the right hand side exactly cancels the two bosonic
determinants. The remaining integral over 
$N-1$ eigenvalues is equal to the quenched partition function which does
not depend on the mass. The only mass dependence is from the weight function, 
$w$, evaluated at the mass $z$ and from the divergence (\ref{ediv}).
We thus find
\be
{\cal Z}_N^{N_f=0,n=-1}(z,z^*;\mu)  \sim \frac {w(z,z^*;\mu)}{|z|^2} ,
\ee
which is exactly the result (\ref{Znm1}) given in section IV.

The absolute value squared pole responsible for attracting the eigenvalue
$z_N$ to the mass $z$ appears for all nonzero values of $\mu$. Hence the
$n=-1$ theory must be regulated for all nonzero values of $\mu$. The
  transition to $\mu \to 0$ is discontinuous because if $\mu = 0$ the 
probability to find an eigenvalue near $z$ is zero unless $z$ 
is on the imaginary axis.

\section{Summary and Discussion}

We have computed the microscopic limit (also known as the $\epsilon$-limit) 
of the QCD partition function 
at nonzero chemical potential  with one bosonic quark. For the computation we
made use of the random matrix representation of the partition function.
Contrary to the partition function with one fermionic
flavor this partition function depends on the chemical potential. In
particular it has been shown that a phase transition takes place at 
$\mu=m_\pi/2$. The theory with a pair of conjugate bosonic flavors also
behaves differently from its fermionic counterpart. While a phase transition
signaled by Bose-Einstein condensation of pions takes place in the fermionic
theory, its bosonic variant always remains in a $\mu$ dependent phase. The
main differences between fermionic and bosonic partition functions 
have been summarized in table 1.

Bosonic partition functions as studied in this paper also appear
in the closely related Hatano-Nelson model \cite{HN}. This is a model
for a disordered system in an  imaginary vector potential. The random
matrix limit of this model
is a non-Hermitian random matrix model  
just like that in (\ref{ZNfNb}) except for the chiral block structure in
(\ref{dnew}). The partition function with $N_f=-1$ can again be expressed 
as the Cauchy transform of the orthogonal polynomials. However, because
this model
is technically simpler than the random matrix model for the QCD partition
function, this partition function can also be evaluated directly by means
of standard random matrix techniques. The agreement of  both approaches
\cite{SV-HN} confirms our understanding of the Cauchy transform representation 
of the partition function. 
Moreover, as in the chiral case, we find that the partition
function of the  Hatano-Nelson model with $N_f=-1$ depends on the chemical
potential and  a phase transition appears in the thermodynamic
limit.

These results generalize to theories with more fermionic and bosonic flavors,  
and we expect similar surprises for 2 color QCD and for QCD with
quarks in the adjoint representation.   

Our results might be relevant for a better understanding of the sign problem
in QCD at nonzero baryon chemical potential. The expectation value of
the phase of the fermion determinant is given by
\be
\left\langle\exp(2i\theta)\right\rangle 
&= &\left\langle\frac{\det(D+\mu\gamma_0+m)}{\det(D+\mu\gamma_0+m)^*} 
\right\rangle \nn\\  
&= &\left\langle\frac{{\det}^2(D+\mu\gamma_0+m)}
{\det(D+\mu\gamma_0+m)^*\det(D+\mu\gamma_0+m)} 
\right\rangle .
\ee
It thus corresponds to a partition function with a bosonic quark and
its conjugate and two fermionic quarks. If the same arguments apply
as in the case of the partition function
with one bosonic flavor the vacuum expectation value of $\exp(2i\theta)$ 
will be nonanalytic at $\mu = m_\pi/2$, in the quenched as well as in 
the unquenched theory. We speculate that this nonanalyticity persist 
at nonzero temperatures and that beyond this point the severity of the 
sign problem increases drastically. 

\vspace{0.5cm}

\noindent
{\bf Note added in proof:} The expectation that
$\left\langle\exp(2i\theta)\right\rangle$ is nonanalytic at $\mu = m_\pi/2$,
in the quenched as well as in the unquenched theory has now been verified
\cite{phaseletter}.

\vspace{1cm}

\noindent
{\bf Acknowledgments:} 
We wish to thank Gernot Akemann, James Osborn, Dennis Dietrich, Dominique
Toublan, Poul Henrik Damgaard  and Frank Wilczek
for valuable discussions. 
Gernot Akemann is thanked for pointing out a mistake in an earlier version 
of this paper.
This work was
supported in part by U.S. DOE Grant No. DE-FG-88ER40388. KS was supported by
the Carslberg Foundation.    

\newpage

\renewcommand{\thesection}{Appendix \Alph{section}}
\setcounter{section}{0}

\section{Calculation of the Particle Spectrum of the 
Chiral Lagrangian for a Pair of Bosonic Quarks}

In this appendix we calculate the particle spectrum of the chiral Lagrangian
(\ref{lag-bcb}) for QCD with a pair of conjugate bosonic quarks. We only
work out in detail the case that $z =x$ is real and $\epsilon = 0$.

The
quark fields in the Goldstone manifold $Gl(2)/U(2)$ can be 
parameterized as
\be
Q =
\mat \cosh s e^{r + t} & \sinh s e^{i\theta + t}  \\
\sinh s e^{-i\theta + t}  &\cosh s e^{- r + t}\emat
\ \ \ \ \ 
Q^{-1} =
\mat \cosh s e^{-r - t} & -\sinh s e^{i\theta - t}  \\
-\sinh s e^{-i\theta - t}  &\cosh s e^{r - t}\emat.
\ee
For the covariant derivatives we find
\be
\nabla_0 Q &\equiv& \del_0 Q - \mu\{Q,B\}\nn \\ 
 &=& \mat  e^{r + t}[ \del_0 \cosh s + \cosh s \del_0( r + t) -2\mu \cosh 
s ]
&e^{i\theta+t}[\del_0 \sinh s + \sinh s\del_0(i\theta + t)]\\ 
e^{-i\theta+t}[\del_0 \sinh s +\sinh s\del_0(-i\theta +t)] 
 &  e^{-r + t}[ \del_0 \cosh s + \cosh s \del_0(-r+t) +2\mu\cosh s ]
\emat
\ee
and
\be
\nabla_0 Q^{-1}  &\equiv& \del_0 Q^{-1} +\mu\{Q^{-1},B\}\nn \\
 &=& \mat  e^{-r - t}[ \del_0 \cosh s - \cosh s \del_0(r+t)+2\mu\cosh s]
&-e^{i\theta-t}[\del_0 \sinh s +\sinh s\del_0(i\theta -t) ]\\ 
-e^{-i\theta-t}[\del_0 \sinh s -\sinh s\del_0(i\theta +t)] 
 &  e^{r - t}[ \del_0 \cosh s + \cosh s \del_0(r-t)-2\mu\cosh s]
\emat
\ee
For the time derivatives in the kinetic term we obtain
\be
{\rm Tr} \nabla_0 Q \nabla_0 Q^{-1} &=& 
-2(\del_0s)^2-2(\del_0t)^2 - 2\cosh^2s (\del_0r)^2 -
2\sinh^2s (\del_0 \theta)^2 +8\mu \cosh^2 s \del_0 r -8 \mu^2 \cosh^2s.
\ee

The mass term is given by
\be
- i \frac 12 \Sigma{\rm Tr} M^T (Q - IQ^{-1}I) 
= - i \Sigma \frac 12 (4 x \sinh s \cosh t \cos \theta
-4y\sinh s\sinh t\sin\theta+4\epsilon\cosh r\cosh s\cosh t)  
\ee
The total Lagrangian is 
\be
{\bf -}\frac{ F^2}4   {\rm Tr} \nabla_\mu  Q \nabla Q^\dagger_\mu 
- i \frac 12 \Sigma {\rm Tr} M^T (Q - IQ^{-1}I)  
\label{lag-ch}
\ee
with $M$ and $I$ defined in (\ref{M-bos}).
Notice that because of the noncompact degrees of freedom, the sign of
the kinetic term is opposite to the usual sign. 

The static part of the Lagrangian for $y=0$ and $\epsilon=0$ is given by
\be
{\cal L}_{\rm static} = 2 F^2 \mu^2\cosh^2 s - i2\Sigma x \sinh s \cosh t
\cos \theta -4\epsilon \cosh r \cosh s \cosh t.
\ee
The minimum is at (note that we can always divide out $\cosh s$, so
there is only one minimum)
\be
\sinh s = i\frac {x \Sigma}{2F^2 \mu^2} =
i\frac{m_\pi^2}{4\mu^2}
\ee
and
\be
t = 0, \quad r = 0, \quad \theta =0.
\ee
We expand the chiral Lagrangian (\ref{lag-ch}) to second order
in the small fluctuations about the saddle point.
The second order terms are given by
\be
&&F^2[-\frac 12(\del_0\delta s)^2 - \frac 12(\del_0\delta t)^2
- \frac 12\cosh^2s (\del_0\delta r )^2 - 
\frac 12 \sinh^2s (\del_0\delta \theta)^2
+4\mu \sinh s \cosh s \delta s \del_0 \delta r
-2\mu^2 (\cosh^2 s +\sinh^2s)(\delta s)^2]\nn \\
&& + i\Sigma x \sinh s [(\delta s)^2  +(\delta t)^2 -(\delta \theta)^2].
\ee
We obtain the uncoupled masses 
\be
m_t^2 &=& -i\frac {2x \Sigma}{F^2}\sinh s 
= \frac{m_\pi^4}{4\mu^2},\\
m_\theta^2 & = & 4\mu^2. 
\ee
The remaining two masses of follow from solving
\be
\det
\mat  E^2  - 4\mu^2\cosh^2s  
& + 4\mu E\sinh s \cosh s  \nn \\
- 4  \mu E\sinh s \cosh s  
&E^2 \cosh^2 s \emat =0
\ee
since $\cosh s\neq 0$ we can divide by $\cosh^2 s$ and obtain
\be
E^2(E^2 - 4\mu^2 \cosh^2 s) +16 \mu^2 E^2\sinh^2 s = 0
\ee
with solutions
\be
E^2 = 0 \quad {\rm or }\quad E^2 = 4\mu^2 \cosh^2 s - 16 \mu^2 \sinh^2 s 
= 4\mu^2 (1 + \frac {3 m_\pi^2}{16\mu^4})
\ee
Note that one mode is massless for all $\mu>0$. 
For $\epsilon \ne 0$ one can show along the same lines that the massless
mode obtains a mass $\sim \sqrt \epsilon$.

\vspace*{1cm}\noindent


\end{document}